\documentclass{ifacconf}

\usepackage{color,graphicx,epstopdf}     
\usepackage{natbib}        
\usepackage{amsmath, amssymb, amsfonts, bm}
\usepackage{mathrsfs}

\newtheorem{theorem}{Theorem}
\newtheorem{definition}{Definition}

\newtheorem{lemma}{Lemma}
\newtheorem{remark}{Remark}

\newtheorem{assumption}{Assumption}

\newcommand{\mcalV}{\mathcal{V}}
\newcommand{\mcalN}{\mathcal{N}}
\newcommand{\mcalE}{\mathcal{E}}

\begin{document}
	\begin{frontmatter}
		
		\title{On the Stability of Networked Nonlinear Negative Imaginary Systems with Applications to Electrical Power Systems}
		
		\thanks[footnoteinfo]{This work was supported by the Australian Research Council under grants DP230102443 and LP210200473.}
		
		\author[First]{Yijun Chen}, \author[First]{Kanghong Shi}, \author[First]{Ian R. Petersen}, \author[First]{Elizabeth L. Ratnam}
		
		\address[First]{The  School of Engineering, The Australian National University, Canberra, Australia, emails: \{yijun.chen, kanghong.shi, ian.petersen, elizabeth.ratnam\}@anu.edu.au.}
		
		\begin{abstract}               
			In the transition to achieving net zero emissions, it has been suggested that a substantial expansion of electric power grids will be necessary to support emerging renewable energy zones. In this paper, we propose employing battery-based feedback control and nonlinear negative imaginary (NI) systems theory to reduce the need for such expansion. By formulating a novel Lur\'e-Postnikov-like Lyapunov function, stability results are presented for the feedback interconnection of two single nonlinear NI systems, while output feedback consensus results are established for the feedback interconnection of two networked nonlinear NI systems based on a network topology. This theoretical framework underpins our design of battery-based control in power transmission systems. We demonstrate that the power grid can be gradually transitioned into the proposed NI systems, one transmission line at a time. 
		\end{abstract}
		
		
	\end{frontmatter}
	
	\section{introduction}
	To enable the energy transition towards net zero power systems, it has been proposed that a significant expansion of the transmission grid will be necessary to accommodate the emergence of renewable energy zones \citep{AEMO_TE}. In this paper, we propose employing battery-based feedback control and nonlinear negative imaginary (NI) systems theory \citep{petersen2010feedback,lanzon2008stability,shi2023output} to mitigate the need for such expansion, thereby facilitating fuller utilization of existing grid infrastructure.
	
	In recent years, advancements in battery technologies have led to the widespread adoption of rechargeable batteries in electric vehicles, large grid storage batteries, and domestic solar-powered batteries \citep{tran2019efficient,borenstein2022s}. Apart from supporting local power management, this transformative development also presents the possibility for active participation in frequency and power flow regulation, as well as ensuring transient stability within power systems. In pursuit of this goal, this paper provides a systematic method to design feedback controllers based on the use of battery-based actuators to synchronize the grid frequency across the transmission network, regulate power flows, and ensure bulk system transient stability.
	
	Real-time angle measurements hold potential for future power systems, as discussed by \citep{vivsic2020synchronous}. NI systems theory can be utilized to guarantee system stability, enhance system robustness, and address consensus problems. Accordingly, the authors in \citep{chen2023design} propose angle feedback linearization control using linear NI systems theory to enhance the transient stability of power transmission systems. However, the method proposed in \citep{chen2023design} directly cancels out the nonlinear properties of power flow along transmission lines. In contrast, the work of \citep{chen2023nonlinear} proposed nonlinear angle feedback control for a single-machine infinite bus system. Although the method in \citep{chen2023nonlinear} uses nonlinear techniques, the authors consider the more simple case of a single generator bus connected to an infinite bus, ruling out its application to the more realistic case of an interconnected transmission network. 
	
	In this paper, we propose a networked control framework for power transmission systems that retains the nonlinear properties of power flow, which is specifically designed to accommodate an interconnected transmission network. From a technical standpoint, we constructed a novel candidate Lyapunov function, fashioned in the Lur\'e-Postnikov form \citep{haddad2008nonlinear,hill1989lyapunov,hill1982stability,bergen1981structure}, intended for stability proofs for the feedback interconnection of two single nonlinear NI systems and output consensus proofs for the feedback interconnection of two networked nonlinear NI systems based on the network structure. Our contributions to the area of power transmission systems are two-fold: 1) in the absence of controllers, we show that existing power transmission systems are resilient --- that is, resilient power transmission systems synchronize bus frequencies and regulate power flows when initial angle deviations are within a suitable domain; 2) we present a networked control framework using real-time angle sensors and large-scale batteries to enhance the transient stability of power transmission systems, where the control can be realized one transmission line at a time. In order to achieve these results, we extend the networked nonlinear NI theory of \citep{shi2023output} to allow for nonlinear direct feedthrough terms and Lur\'e-Postnikov type Lyapunov functions. 
	
	This paper is organized as follows. Section \ref{sec:stability_single} provides definitions for NI systems, and stability results for the feedback interconnection of two single nonlinear NI systems. In Section~\ref{sec:networked_case}, a networked setting is considered, and output feedback consensus is proved for two networked nonlinear NI systems. In Section~\ref{sec:power_application}, we present an application to power transmission systems. 

	\section{An Initial Stability Result}\label{sec:stability_single}
	In this section, we present definitions for nonlinear NI systems and include our stability results for the feedback interconnection of two single nonlinear NI systems.
	
	\subsection{Definitions of Nonlinear NI Systems}
	Consider a multiple-input multiple-output (MIMO) nonlinear system with the following state-space model:
	\begin{subequations}
		\label{sys:mimo_sp}
		\begin{align}
			\dot{x} &= f(x,u), \label{eq:mimo_state}\\
			y & = h(x)+g(u), \label{eq:mimo_output}
		\end{align}
	\end{subequations}
	where 
	$x \in \mathbb{R}^{n}$ is the state, $u \in \mathbb{R}^{m}$ is the input, $y \in \mathbb{R}^{m}$ is the output, $f:\mathbb{R}^{n} \times \mathbb{R}^{m} \to \mathbb{R}^{n}$ is a Lipschitz continuous function, and $h:\mathbb{R}^{n} \to \mathbb{R}^{m}$ is a class C$^{1}$ function. We impose Assumption~\ref{apt:g} on the input function $g(u)$.  As a special case, we consider a static system $y = g(u)$ of the system~\eqref{sys:mimo_sp}. In general, nonlinear controllers are dynamic unless otherwise stated for the special static case.
	\begin{assumption}\label{apt:g} 
		The input function $g(u)$ is independent in each input channel, such that
		\begin{equation}\label{eq:independent_input}
			g(u) = [g^{1}(u^{1}), \dots, g^{m}(u^{m})]^{\top},
		\end{equation}
		where each $g^{k}(u^{k})$ is a class $C^{1}$ function with the superscript $k \in \{1,2,\dots,m\}$ representing the $k$th element of the input $u$. Moreover, $g(0) = 0.$
	\end{assumption}

	
	In this paper, nonlinear generalizations of standard properties for linear systems are assumed as done in \citep{shi2023output}. The following Assumption~\ref{apt:state_output_consistency} is an observability assumption, while Assumption~\ref{apt:input_state_consistency} requires all system inputs to have an effect on the system dynamics.
	\begin{assumption}\label{apt:state_output_consistency}
		Over any time interval $[t_a,t_b]$ where $t_b>t_a$, $h(x)$ remains constant if and only if $x$ remains constant; i.e., $\dot{h}(x) \equiv 0\iff \dot x \equiv 0$. Moreover, $h(x)  \equiv 0 \iff x\equiv 0$. 
		
	\end{assumption}
	
	\begin{assumption}\label{apt:input_state_consistency}
		Over any time interval $[t_a,t_b]$ where $t_b>t_a$, $x$ remains constant only if $u$ remains constant; i.e., $x\equiv\overline x {\implies} u\equiv\overline u $. Moreover, $x\equiv 0 {\implies} u\equiv 0$.
	\end{assumption}

	Next, we define the negative imaginary (NI) property, and we include the definition of output strictly negative imaginary (OSNI) property ---  tailored for nonlinear MIMO systems. 
	\begin{definition}
		\label{def:mimo_ni}
		The system~\eqref{sys:mimo_sp} is said to be 
		NI if there exists a positive semidefinite storage function $V:\mathbb{R}^{n} \to \mathbb{R}$ of class C$^{1}$ such that for any locally integrable input $u$ and solution $x$ to (\ref{eq:mimo_state}), then
		\begin{align}\label{eq:general_dissipation_inequality}
			\dot{V}(x) &\leq u^{\top}\dot{h}(x),
		\end{align}
  	for all $t \geq 0$.
	\end{definition}
	
	\begin{definition}
		\label{def:mimo_osni}
		The system~\eqref{sys:mimo_sp} is said to be 
		OSNI if there exists a positive semidefinite storage function $V:\mathbb{R}^{n} \to \mathbb{R}$ of class C$^{1}$ and a scalar $\epsilon > 0$ such that for any locally integrable input $u$ and solution $x$ to (\ref{eq:mimo_state}), then\begin{align}\label{eq:strict_dissipation_inequality}
			\dot{V}(x) &\leq u^{\top}\dot{h}(x) - \epsilon\|\dot{h}(x)\|^{2},
		\end{align}
 for all $t \geq 0$.
		Here, $\epsilon$ measures the degree of output strictness.
	\end{definition}
	
	\subsection{Stability of NI Systems}

	
	In what follows, we present some initial stability results for the positive feedback interconnection of an NI system and an OSNI system.
	
	Consider a nonlinear system $H_{p}$: 
	\begin{subequations}
		\label{sys:h1_sp}
		\begin{align}
			H_{p} : \quad	\dot{x}_{p} &= f_{p}(x_{p},u_{p}), \label{eq:h1_state}\\
			y_{p} & = h_{p}(x_{p}), \label{eq:h1_output}
		\end{align}
	\end{subequations}
	where $x_{p} \in \mathbb{R}^{n_{p}}$ is the state, $u_{p} \in \mathbb{R}^{m}$ is the input, $y_{p} \in \mathbb{R}^{m}$ is the output, $f_{p}:\mathbb{R}^{n_{p}} \times \mathbb{R}^{m} \to \mathbb{R}^{n_{p}}$ is a Lipschitz continuous function, and $h_{p}:\mathbb{R}^{n_{p}} \to \mathbb{R}^{m}$ is a class C$^{1}$ function. The subscript ``p'' implies that this system plays the role of a plant. Also, we assume $f_p(0,0)=0$ and $h_p(0)=0$.
	
	\begin{assumption}\label{apt:positive_uh}
		For a system $H_{p}$ with a constant input $\overline{u}_{p}$ which results in a  constant state $\overline{x}_{p}$ and a constant output $\overline{y}_{p}$, then $\overline{u}_{p}^{\top}\overline{y}_{p} \geq 0.$ 
	\end{assumption}
	
	Also consider a  nonlinear system $H_{c}$: 
	\begin{subequations}
		\label{sys:h2_sp}
		\begin{align}
			H_{c} : \quad	\dot{x}_{c} &= f_{c}(x_{c},u_{c}), \label{eq:h2_state}\\
			y_{c} & = h_{c}(x_{c}) + g_{c}(u_{c}), \label{eq:h2_output}
		\end{align}
	\end{subequations}
	where $x_{c} \in \mathbb{R}^{n_{c}}$ is the state, $u_{c} \in \mathbb{R}^{m}$ is the input, $y_{c} \in \mathbb{R}^{m}$ is the output, $f_{c}:\mathbb{R}^{n_{c}} \times \mathbb{R}^{m} \to \mathbb{R}^{n_{c}}$ is a Lipschitz continuous function, and $h_{c}:\mathbb{R}^{n_{c}} \to \mathbb{R}^{m}$ is a class C$^{1}$ function. Assumption~\ref{apt:g} is assumed for the input function $g_{c}(u_{c})$. The subscript ``c'' indicates that this system serves as a controller. We allow for the special case of a static system $y_{c} = g_{c}(u_{c})$, that is,  a special case of the system~\eqref{sys:h2_sp} which is NI according to Definition~\ref{def:mimo_ni} with storage function $V_c=0$. Accordingly, we assume $f_c(0,0)=0$ and $h_c(0)=0$.
	
	\begin{assumption}\label{apt:negative_uy}
		For a system $H_{c}$ with a constant input $\overline{u}_{c}$ which results in a constant output $\overline{y}_{c}$, then $\overline{u}_{c}^{\top}\overline{y}_{c}\leq -\gamma_{c}\| \overline{u}_{c}\|^{2}$ with $\gamma_{c} >0.$ 
	\end{assumption}

	\begin{figure}[h]
		\centering
		\includegraphics[width=0.5\linewidth]{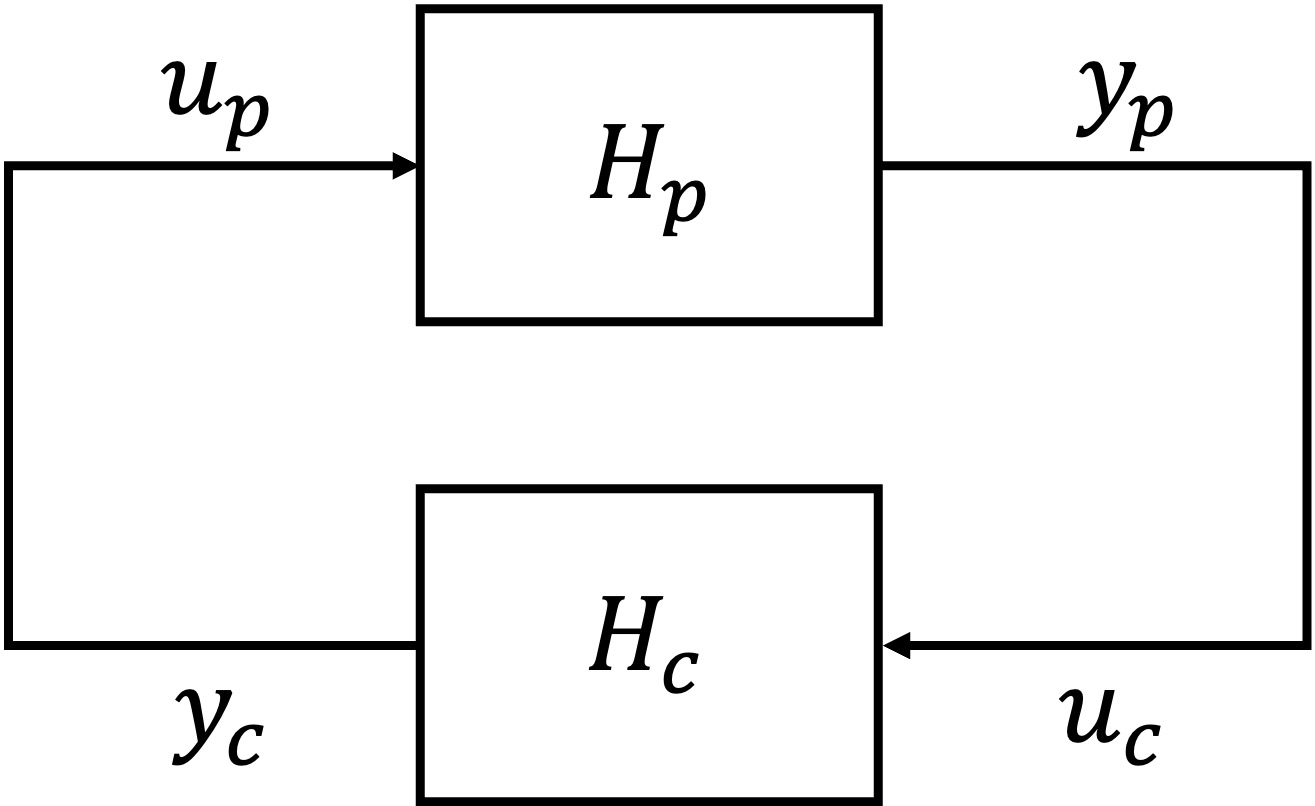}
		\caption{The feedback interconnection of  the nonlinear  system $H_{p}$ and the nonlinear system $H_{c}$.}
		\label{fig:single_loop}
	\end{figure}
	
	Further, consider the feedback interconnection of the nonlinear system $H_{p}$ and the nonlinear system $H_{c}$ as shown in Fig.~\ref{fig:single_loop}. The relationship between the inputs  and the outputs are described by 
	\begin{subequations}\label{eq:inputs_outputs}
		\begin{align}
			&y_{p} \equiv h_{p}(x_{p})   \equiv u_{c}, \label{eq:y_p_u_c}\\
			&y_{c} \equiv h_{c}(x_{c}) + g_{c}(u_{c})\equiv u_{p}. \label{eq:y_c_u_p}
		\end{align}
	\end{subequations}
According to (\ref{eq:y_p_u_c}) and (\ref{eq:y_c_u_p}),	we obtain the following equation
	\begin{equation}\label{eq:equal_mutiplication}
		u_{p}^{\top}y_{p} \equiv u_{c}^{\top}y_{c}.
	\end{equation}
	
	For the feedback system $(H_{p},H_{c})$ in Fig.~\ref{fig:single_loop}, we define a candidate of Lur\'e-Postnikov-like Lyapunov function \citep{haddad2008nonlinear} as
	\begin{align}
		\label{eq:fb_storage}
		W(x_{p},x_{c}) = & \  V_{p}(x_{p}) + V_{c}(x_{c}) - h_{p}(x_{p})^{\top}h_{c}(x_{c}) \notag \\
		& - \sum_{k=1}^{k=m} \int_{0}^{h_{p}^{k}(x_{p})}  g_{c}^{k}(\xi^{k})d\xi^{k}.
	\end{align}
	
	\begin{assumption}
		\label{apt:local_domain}
	There exists an open domain $\mathcal{D} \subset \mathbb{R}^{n_{p}} \times \mathbb{R}^{n_{c}}$ containing the origin such that the candidate Lyapunov function~\eqref{eq:fb_storage} is positive definite on $\mathcal D$.
	\end{assumption}
	\medskip

	The following theorems establish stability results for the feedback  system $(H_{p},H_{c})$.
	\medskip

		\begin{theorem}
			\label{thm:OSNI_plant_static/NI_controller}
			Consider a nonlinear OSNI plant $H_{p}$ satisfying Assumptions~\ref{apt:state_output_consistency},~\ref{apt:input_state_consistency},  and \ref{apt:positive_uh}. Also, consider a nonlinear controller  $H_{c}$,  either being a nonlinear static controller or a nonlinear dynamic NI controller. Suppose Assumptions~\ref{apt:g} and \ref{apt:negative_uy} hold for the nonlinear static controller, and Assumptions~\ref{apt:g}, \ref{apt:state_output_consistency}, and \ref{apt:negative_uy} hold for the nonlinear dynamic NI controller. Further, consider the feedback system $(H_{p},H_{c})$ as depicted in Fig~\ref{fig:single_loop}. Also, suppose Assumption~\ref{apt:local_domain} holds for the feedback system $(H_{p},H_{c})$. Then,  the feedback  system $(H_{p},H_{c})$ is locally asymptotically stable.
	\end{theorem} 
	{\it  Proof.} First, according to Assumption~\ref{apt:local_domain}, the candidate Lyapunov function~\eqref{eq:fb_storage} is positive definite in the domain $\mathcal{D}$. Second, since both the static controller and the dynamic NI controller are NI, the time derivative of the candidate Lyapunov function~\eqref{eq:fb_storage} is analyzed as follows:
	\begin{align}\label{eq:dynamic_dot_W}
		\dot{W} 
		\overset{Eqs.~\eqref{eq:general_dissipation_inequality}, \eqref{eq:strict_dissipation_inequality}}{\leq} &  (u_{p} -h_{c}(x_{c}) - g_{c}(h_{p}(x_{p})))^{\top}\dot h_{p}(x_{p})  \notag\\
		& - \epsilon \| \dot h_{p}(x_{p})\|^{2}  + (u_{c} - h_{p}(x_{p}))^{\top}\dot{h}_{c}(x_{c}) \notag\\
		\overset{Eq.~\eqref{eq:inputs_outputs}}{\leq} &    -\epsilon_{p} \| \dot{h}_{p}(x_{p}) \|^{2}   \leq  0.
	\end{align}
	
	Thus, in both the cases of a static controller and a dynamic controller,  the feedback system $(H_{p},H_{c})$ is at least locally stable in the sense of Lyapunov. 
	
	Equation~\eqref{eq:dynamic_dot_W} implies that $\dot{W}$ can remain zero only if $\dot{h}_{p}(x_{p})$ remains zero. 
		This implies that the OSNI plant $H_{p}$ reaches steady state such that 
				$\dot{h}_{p}(x_{p})  \equiv 0 \overset{\text{A}2}{\implies} \overline{y}_{p} \overset{\text{A}2}{\implies} \overline{x}_{p}  \overset{\text{A}3}{\implies} \overline{u}_{p} \text{ and }
				 \overline{u}_{p}^{\top}\overline{y}_{p}  \overset{\text{A}4}{\geq} 0.$ 
		Then, according to Eq.~\eqref{eq:inputs_outputs} and Eq.~\eqref{eq:h2_output}, the nonlinear controller $H_{c}$ also reaches a steady state such that
			$\overline{u}_{c}^{\top}\overline{y}_{c} 
			\overset{\text{A}5}{\leq} -\gamma_{c} \|\overline{u}_{c} \|^{2}.$ 
		Thus,  $\overline{u}_{c}$ can only be zero, which yields
		$\overline{y}_{p} \overset{Eq.~\eqref{eq:y_p_u_c}}{\equiv} \overline{u}_{c} \equiv 0.$
		Then, for the system $H_{p}$, we obtain that
		$\overline{y}_{p} \equiv 0 \overset{\text{A}2}{\implies} \overline{x}_{p} \equiv  0.$ 
		
		Next, consider the cases of a nonlinear static controller and a nonlinear dynamic controller. In the case of a nonlinear static controller, there are no dynamics. Thus, we can directly conclude that $\dot{W}$ cannot remain zero unless $x_{p} = 0$. According to LaSalle's invariance principle, $W(x_{p})$ will keep decreasing until $x_{p} = 0$. In the case of a nonlinear dynamic NI controller, where
		$
		\overline{x}_{p} \equiv 0 \overset{\text{A}2, \text{A}3}{\implies} \overline{y}_{p} \equiv 0 \text{ and } \overline{u}_{p} \equiv 0  \overset{Eq.~\eqref{eq:inputs_outputs}}{\implies} \overline{u}_{c} \equiv 0 \text{ and } \overline{y}_{c} \equiv 0 \overset{Eq.~\eqref{eq:h2_output}}{\implies} \dot{h}_{c}(x_{c}) \equiv 0 \overset{ \text{A}2}{\implies} x_{c} \equiv 0
		$, we observe that $\dot{W}$ cannot remain zero unless $x_{p} = 0$ and $x_{c} = 0$. According to LaSalle's invariance principle, $W(x_{p},x_{c})$ decreases until $x_{p} = 0$ and $x_{c} = 0$. Therefore, in both cases, the feedback system $(H_{p}, H_{c})$ is locally asymptotically stable. The proof is now completed. \hfill $\square$
	\medskip
	
	\begin{theorem}
		\label{thm:NI_plant_OSNI_controller}
		Consider a nonlinear NI plant $H_{p}$ satisfying Assumptions~\ref{apt:state_output_consistency},~\ref{apt:input_state_consistency},  and \ref{apt:positive_uh}. Also, consider a nonlinear OSNI controller  $H_{c}$ satisfying Assumptions~\ref{apt:g},~\ref{apt:state_output_consistency},~\ref{apt:input_state_consistency}, and \ref{apt:negative_uy}. Further, consider the feedback system $(H_{p},H_{c})$ as depicted in Fig~\ref{fig:single_loop}. Suppose Assumption~\ref{apt:local_domain} holds for the feedback system $(H_{p},H_{c})$. Then,  the feedback  system $(H_{p},H_{c})$ is locally asymptotically stable.
	\end{theorem} 
	{\it Proof.} The proof is similar to that for Theorem~\ref{thm:OSNI_plant_static/NI_controller}. \hfill$\square$

	\section{Output Consensus of Networked  NI Systems}\label{sec:networked_case}
	In this section, a network setting is considered, and output consensus results are presented for the feedback interconnection of two networked NI systems.
	
	\subsection{Settings for Networked NI Systems}
	{\bf Network Setting.} In what follows, we consider a connected and undirected network $\mathcal{G} = (\mathcal{V}, \mathcal{E})$, where $\mathcal{V} = \{1, 2, \dots, N\}$ describes the set of $N$ nodes, and $\mathcal{E} = \{e_{1}, e_{2}, \dots, e_{L}\} \subseteq \mathcal{V} \times \mathcal{V}$ represents the set of $L$ edges connecting the nodes. The index set for edges is denoted by $\mathcal{L}=\{1,2,\dots,L\}.$ Each node is associated with an independent nonlinear plant, while each edge is deployed with a nonlinear controller. Each edge takes the outputs of two end nodes as its input, and each node takes the outputs of its connected edges as its input.
	
	If there exists an edge $(i, j) \in \mathcal{E}$ connecting node $i \in \mathcal{V}$ and node $j \in \mathcal{V}$, then nodes $i$ and $j$ are considered neighbors. The neighbors of node $i \in \mathcal{V}$ are indexed in the set $\mathcal{N}(i)$. The edges that contain node $i$ are indexed in the set $\mathcal{E}(i)$.  
	The incidence matrix $\mathbf{Q} \in \mathbb{R}^{N \times L}$ of the network is defined as follows:
	\begin{equation*}
		\mathbf{Q}_{ie} = 
		\begin{cases}
			1, & \text{if } i \text{ is the initial node of edge } e,\\
			-1, & \text{if } i \text{ is the terminal node of edge } e,\\
			0, & \text{if } i \text{ is not connected in edge } e.
		\end{cases}
	\end{equation*}
	It is noted that a fixed representation of edges is chosen, where each $(i, j)$ or $(j, i)$ can only be chosen once, and ``initial/terminal node'' does not refer to a specific orientation.
	
	{\bf Node Plants.} Each node $i \in \mathcal{V}$ is associated with an independent  nonlinear plant $H_{pi}$ described by:
	\begin{subequations}
		\label{sys:networked_h1_sp}
		\begin{align}
			H_{pi} : \quad	\dot{x}_{pi} &= f_{pi}(x_{pi},u_{pi}), \label{eq:networked_h1_state}\\
			y_{pi} & = h_{pi}(x_{pi}), \label{eq:networked_h1_output}
		\end{align}
	\end{subequations}
	where $x_{pi} \in \mathbb{R}^{n_{pi}}$ is the state, $u_{pi} \in \mathbb{R}^{m}$ is the input, $y_{pi} \in \mathbb{R}^{m}$ is the output, $f_{pi}:\mathbb{R}^{n_{pi}} \times \mathbb{R}^{m} \to \mathbb{R}^{n_{pi}}$ is a Lipschitz continuous function, and $h_{pi}:\mathbb{R}^{n_{pi}} \to \mathbb{R}^{m}$ is a class C$^{1}$ function. Also, we assume $f_{pi}(0,0)=0$ and $h_{pi}(0)=0$. For a compact expression, we collect the states, inputs and outputs of all nodes --- as represented by the aggregated state vector $X_{p}=[x_{p1}^{\top}, \dots, x_{pN}^{\top}]^{\top} \in \mathbb{R}^{n_{p}}$ with $n_{p} = \sum_{i = 1}^{N}n_{pi}$, the aggregated input vector $U_{p}=[u_{p1}^{\top}, \dots, u_{pN}^{\top}]^{\top} \in \mathbb{R}^{mN}$, and the aggregated output vector $Y_{p}=[y_{p1}^{\top}, \dots, y_{pN}^{\top}]^{\top}\in \mathbb{R}^{mN}$. We denote the aggregated node plants by $\mathcal{H}_{p}$, which is described by 
	\begin{align*}
		\mathcal{H}_{p}: \dot{X}_{p} = \begin{bmatrix}
			f_{p1}(x_{p1},u_{p1})\\
			\vdots\\
			f_{pN}(x_{pN},u_{pN})
		\end{bmatrix},
		Y_{p} = \begin{bmatrix}
			h_{p1}(x_{p1})\\
			\vdots\\
			h_{pN}(x_{pN})
		\end{bmatrix}.
	\end{align*}
	\medskip
	
	{\bf Edge Controllers.} Each edge $e_{l} \in \mathcal{E}$ with $l \in \mathcal{L}$ is deployed with a  nonlinear controller described by 
	\begin{subequations}
		\label{sys:networked_h2_sp}
		\begin{align}
			H_{cl} : \quad	\dot{x}_{cl} &= f_{cl}(x_{cl},u_{cl}), \label{eq:networked_h2_state}\\
			y_{cl} & = h_{cl}(x_{cl}) + g_{cl}(u_{cl}), \label{eq:networked_h2_output}
		\end{align}
	\end{subequations}
	where $x_{cl} \in \mathbb{R}^{n_{cl}}$ is the state, $u_{cl} \in \mathbb{R}^{m}$ is the input, $y_{cl} \in \mathbb{R}^{m}$ is the output, $f_{cl}:\mathbb{R}^{n_{cl}} \times \mathbb{R}^{m} \to \mathbb{R}^{n_{cl}}$ is a Lipschitz continuous function, and $h_{cl}:\mathbb{R}^{n_{cl}}  \to \mathbb{R}^{m}$ is a class C$^{1}$ function. Also, we assume $f_{cl}(0,0)=0$ and $h_{cl}(0)=0$. Assumption~\ref{apt:g} is assumed for the input functions $g_{cl}(u_{cl}), l \in \mathcal{L}$. We allow for the case of a static system $y_{cl} = g_{cl}(x_{cl}), l \in \mathcal{L}$ as a special case of the system~\eqref{sys:networked_h2_sp}.  For a compact expression, we collect the states, the inputs and the outputs of all edges into the aggregated state vector $X_{c}=[x_{c1}^{\top}, \dots, x_{cL}^{\top}]^{\top} \in \mathbb{R}^{n_{c}}$ with $n_{c} = \sum_{l \in \mathcal{L}}n_{cl}$, the aggregated input vector $U_{c}=[u_{c1}^{\top}, \dots, u_{cL}^{\top}]^{\top} \in \mathbb{R}^{mL}$, and the aggregated output vector 
	\begin{equation}
		\label{eq:output_vector}
		Y_{c}=[y_{c1}^{\top}, \dots, y_{cL}^{\top}]^{\top} = \Pi_{cx}(X_{c}) + \Pi_{cu}(U_{c}) \in \mathbb{R}^{mL}, 
	\end{equation}
	where 
	\begin{subequations}
		\begin{align}
			\Pi_{cx}(X_{c}) = [h_{c1}(x_{c1})^{\top}, \dots, h_{cL}(x_{cL})^{\top}]^{\top} \in \mathbb{R}^{mL},\\
			\Pi_{cu}(U_{c}) = [g_{c1}(u_{c1})^{\top}, \dots, g_{cL}(u_{cL})^{\top}]^{\top} \in \mathbb{R}^{mL}.
		\end{align}
	\end{subequations}
	We denote the aggregated nonlinear controllers by $\mathcal{H}_{c}$, which are described by 
	\begin{align*}
		\mathcal{H}_{c}: \dot{X}_{c} = \begin{bmatrix}
			f_{c1}(x_{c1},u_{c1})\\
			\vdots\\
			f_{cL}(x_{cL},u_{cL})
		\end{bmatrix},
		Y_{c} = \begin{bmatrix}
			h_{c1}(x_{c1}) + g_{c1}(u_{c1})\\
			\vdots\\ 
			h_{cL}(x_{cL}) + g_{cL}(u_{cL})
		\end{bmatrix}.
	\end{align*}
	
	\medskip
	
	{\bf Output Feedback Control Framework.} The objective of our control problem is to achieve output consensus for each node in the network. We now define local output consensus. 
	\begin{definition}[Local Output Consensus]\label{def:output_consensus}
		\ \\
		A distributed output feedback control law achieves local output feedback consensus for a networked system if there exists an open domain $\mathcal{D}_{c} \subset \mathbb{R}^{n_{p}\times n_{c}}$ containing the origin such that $\lim_{t \to \infty}\|y_{pi}(t) - y_{pj}(t)\| = 0,$ for all $ i,j \in \mathcal{V},$ for all initial conditions $(X_{p}(0),X_{c}(0)) \in \mathcal{D}_{c}.$ 
	\end{definition}
	
	\medskip

\begin{figure}[bt]
		\centering
		\includegraphics[width=0.9\linewidth]{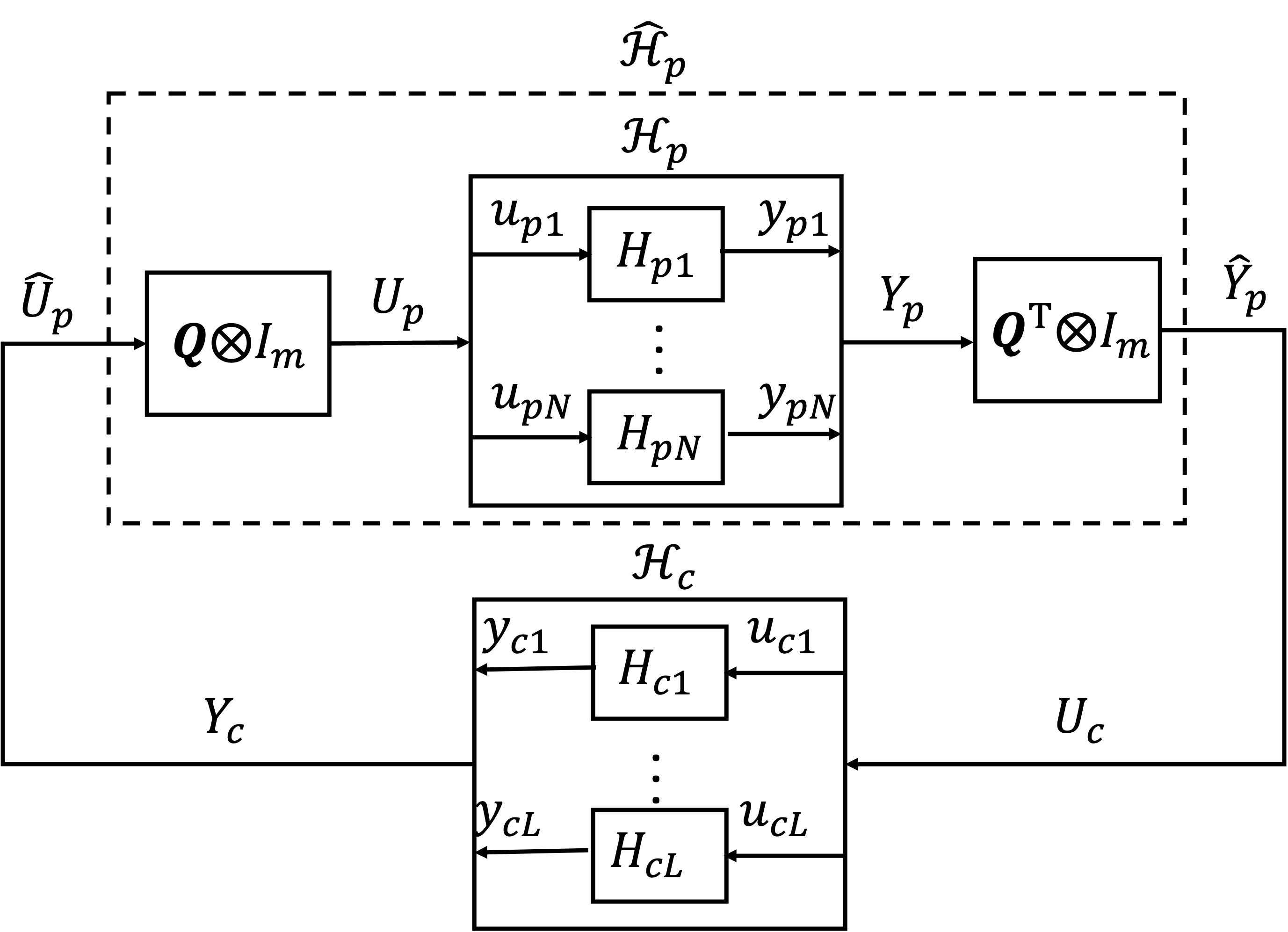}
		\caption{The feedback interconnection of nonlinear plants $\mathcal{H}_{p}$ and nonlinear edge controllers $\mathcal{H}_{c}$ based on  the incidence matrix $\mathbf{Q}$.}
		\label{fig:networked_sys}
	\end{figure}

	As depicted in Fig.~\ref{fig:networked_sys},  a distributed output feedback control framework naturally arises based on  the incidence matrix $\mathbf{Q}$. We denote the networked node plants by $\widehat{\mathcal{H}}_{p} = (\mathbf{Q}^{\top} \otimes I_{m})\mathcal{H}_{p}(\mathbf{Q}\otimes I_{m})$. We further denote the feedback interconnection of  the networked node plants $\widehat{\mathcal{H}}_{p}$ and the aggregated edge controllers $\mathcal{H}_{c}$ by $(\widehat{\mathcal{H}}_{p}, \mathcal{H}_{c})$. The relationship between the inputs and the outputs of the feedback system $(\widehat{\mathcal{H}}_{p}, \mathcal{H}_{c})$  in Fig.~\ref{fig:networked_sys} are described by 
	\begin{subequations}\label{eq:network_input_output}
		\begin{align}
			& \widehat{U}_{p} \equiv Y_{c} \equiv \Pi_{cx}+\Pi_{cu}, \label{eq:U_hat_p_Y_c}\\
			& \widehat{Y}_{p} \equiv U_{c}, \label{eq:Y_hat_p_U_c}\\
			& U_{p} = (\mathbf{Q} \otimes I_{m})\widehat{U}_{p}, \label{eq:U_p_U_hat_p}
			\\
			& \widehat{Y}_{p} \equiv (\mathbf{Q}^{\top}\otimes I_{m}) Y_{p}.  \label{eq:Y_hat_p_Y_p}
		\end{align}
	\end{subequations}
	In a distributed manner, each edge controller $l \in \mathcal{L}$ takes the difference between the outputs of the neighbouring nodes $i$ and $j$ as its input,
	$u_{cl} = \sum_{k = 1}^{N}q_{kl} y_{pk} = y_{pi} - y_{pj},$
	where $q_{kl}$ represents the $k$th element in the $l$th column of the incidence matrix $\mathbf{Q}$, and the node $i$ and the node $j$ are the initial node and the terminal node of the edge $e_{l}$, respectively. 
	Each node plant $i \in \mathcal{V}$ takes the sum of the  outputs from all its connected edge controllers as its input,
	$u_{pi} =  \sum_{l = 1}^{ L} q_{il}y_{cl},$
	where $q_{il}$ is the $l$th element in the $i$th row of the incidence matrix $\mathbf{Q}.$ 

	


	\subsection{Main Results}
	
	{\bf NI/OSNI Property Preservation.} Let the storage function for each node plant $H_{pi}, i \in \mathcal{V}$ be denoted by $V_{pi}$ and let the storage function for each edge controller $H_{cl}, l \in \mathcal{L}$ be denoted by  $V_{cl}$, respectively. The storage functions for the aggregated node plants $\mathcal{H}_{p}$ and the aggregated edge controller $\mathcal{H}_{c}$ are chosen as $V_{p} = \sum_{i \in \mathcal{V}}V_{pi}$ and $V_{c} = \sum_{l \in \mathcal{L}} V_{cl}$, respectively. For the networked node plants $\widehat{\mathcal{H}}_{p}$, the storage function is chosen as the same for the aggregated node plants $\mathcal{H}_{p}$; i.e., $\widehat{V}_{p} = V_{p} = \sum_{i \in \mathcal{V}}V_{pi}.$
	
	In the following two lemmas, we show that aggregation preserves the NI and OSNI properties.
	
	\begin{lemma}\label{lemma:controller_NI/OSNI_preservation}
		If each edge controller $H_{cl}, l \in \mathcal{L}$ is a nonlinear NI (OSNI) system, then the aggregated edge controllers $\mathcal{H}_{c}$ is also a nonlinear NI (OSNI) system.
	\end{lemma}
	
	{\noindent \it Proof.} (1) NI Property Preservation. According to Definition~\ref{def:mimo_ni}, each edge controller $H_{cl}, l \in \mathcal{L}$ has  a positive semidefinite storage function $V_{cl}(x_{cl})$  such that  	for all $t \geq 0$,
	$
	\dot{V}_{cl}(x_{cl}) \leq u_{cl}^{\top}\dot{h}_{cl}(x_{cl}).
	$
	Next, we analyze the time derivative of $V_{c}$:
	\begin{equation}\label{eq:NI_dot_V_c}
		\dot{V}_{c} = \sum_{l \in \mathcal{L}} \dot{V}_{cl}(x_{cl}) \leq U_{c}^{\top}\dot{\Pi}_{cx}.
	\end{equation}
	Therefore,  the aggregated edge controllers $\mathcal{H}_{c}$ also satisfies Definition~\ref{def:mimo_ni}.
	
	(2) OSNI Property Preservation: According to Definition~\ref{def:mimo_osni}, each edge controller $H_{cl}, l \in \mathcal{L}$ has  a positive semidefinite storage function $V_{cl}(x_{cl})$ and a scalar $\epsilon_{l}$ such that  	for all $t \geq 0$,
	$
	\dot{V}_{cl}(x_{cl}) \leq u_{cl}^{\top}\dot{h}_{cl}(x_{cl}) - \epsilon_{cl}\|\dot h_{cl}(x_{cl}) \|^{2}.
	$
	Then, we analyze the time derivative of $V_{c}$:
	\begin{equation}\label{eq:OSNI_dot_V_c}
		\dot{V}_{c} = \sum_{l \in \mathcal{L}} \dot{V}_{cl}(x_{cl}) \leq U_{c}^{\top}\dot{\Pi}_{cx} - \epsilon_{c\min}\|\dot \Pi_{cx}\|^{2},
	\end{equation}
	where $\epsilon_{c\min} = \min\{\epsilon_{c1},\dots, \epsilon_{cL}\} > 0.$
	Therefore,  the aggregated edge controllers $\mathcal{H}_{c}$ also satisfies Definition~\ref{def:mimo_osni}. The proof is now completed.  \hfill$\square$
	
	\begin{lemma}\label{lemma:plant_NI/OSNI_preservation}
		If each node plant $H_{pi}, i \in \mathcal{V}$ is a nonlinear NI (OSNI) system, then the aggregated node plants $\mathcal{H}_{p}$ are also a nonlinear NI (OSNI) system.
	\end{lemma}
	{\it Proof.} The form of the node plants is a special case of the form of the edge controller where the output is only determined by the state. Therefore, the proof of Lemma~\ref{lemma:controller_NI/OSNI_preservation} applies for Lemma~\ref{lemma:plant_NI/OSNI_preservation}. The proof is now completed.  \hfill$\square$
	
	In the following lemma, we present a preliminary result that is needed for the proof of output consensus.
	\begin{lemma}\label{lemma:post-processed}
		If each node plant $H_{pi}, i \in \mathcal{V}$ is a nonlinear NI system, then the storage function for the networked node plants $\widehat{\mathcal{H}}_{p}$ is such that 
		\begin{equation}\label{eq:V_hat_NI}
			\dot{\widehat{V}}_{p} \leq \widehat{U}_{p}^{\top}\dot{\widehat{Y}}_{p}.
		\end{equation} 
		Also, if each node plant $H_{pi}, i \in \mathcal{V}$ is a nonlinear OSNI system, then the storage function for the networked node plants $\widehat{\mathcal{H}}_{p}$ are such that 
		\begin{equation}\label{eq:V_hat_OSNI}
			\dot{\widehat{V}}_{p}  \leq \widehat{U}_{p}^{\top}\dot{\widehat{Y}}_{p} - \epsilon_{p\min}\|\dot{Y}_{p}\|^{2},
		\end{equation}
		where $\epsilon_{p\min} = \min\{\epsilon_{p1},\dots,\epsilon_{pN}\}$.
	\end{lemma}
	{\it Proof. } (1)  NI case.
	If each node plant $H_{pi}, i \in \mathcal{V}$ is NI, then it has a positive semidefinite storage function $V_{pi}(x_{pi})$ such that 	for all $t \geq 0$,
	$
	\dot{V}_{pi}(x_{pi}) \leq u_{pi}^{\top}\dot{h}_{pi}(x_{pi}).
	$	In compact form, we have
	$
	\dot{\widehat{V}}_{p} = \sum_{i \in \mathcal{V}} \dot{V}_{pi}(x_{pi}) \leq U_{p}^{\top}\dot{Y}_{p}.
	$
	According to the input-output relation~\eqref{eq:network_input_output}, we have $U_{p}^{\top}\dot{Y}_{p} = \widehat{U}_{p}^{\top}(\mathbf{Q}^{\top} \otimes I_{m})\dot{Y}_{p} =  \widehat{U}_{p}^{\top}\dot{\widehat{Y}}_{p}.$ Therefore, we obtain $\dot{\widehat{V}}_{p} \leq \widehat{U}_{p}^{\top}\dot{\widehat{Y}}_{p}$. 
	
	(2) OSNI case. If each node plant $H_{pi}, i \in \mathcal{V}$ is OSNI, then it has a positive semidefinite storage function $V_{pi}(x_{pi})$ such that 	for all $t \geq 0$,
	$
	\dot{V}_{pi}(x_{pi}) \leq u_{pi}^{\top}\dot{h}_{pi}(x_{pi}) - \epsilon_{pi}\|\dot h_{pi}(x_{pi})\|^{2}.
	$
	Similar to the proof in (1), we obtain the compact form
	$
	\dot{\widehat{V}}_{p} = \sum_{i \in \mathcal{V}} \dot{V}_{pi}(x_{pi}) \leq \widehat{U}_{p}^{\top}\dot{\widehat{Y}}_{p} - \epsilon_{p\min}\|\dot{Y}_{p}\|^{2},
	$
	where $\epsilon_{p\min} = \min\{\epsilon_{p1},\dots,\epsilon_{pN}\}$. The proof is now completed.  \hfill$\square$
	
	{\bf Output Consensus.} For the feedback system $(\widehat{\mathcal{H}}_{p}, \mathcal{H}_{c})$, a candidate Lyapunov function is selected as 
	\begin{align}
		\label{eq:networked_Lyapunov}
		\widehat{W} = & \sum_{i\in \mathcal{V}}V_{pi}(x_{pi}) + 	\sum_{l\in \mathcal{L}}V_{cl}(x_{cl}) - \widehat{Y}_{p}^{\top}\Pi_{cx} \notag\\
		& - \sum_{k=1}^{mL}\int_{0}^{\widehat{Y}_{p}^{k}}\Pi_{cu}^{k}(\xi^{k})d\xi^{k}.
	\end{align}
	
	\begin{assumption}\label{apt:networked_local_domain}
		There exists an open domain $\mathcal{D} \subset \mathbb{R}^{n_{p}} \times \mathbb{R}^{n_{c}}$ such that the candidate Lyapunov function~\eqref{eq:networked_Lyapunov} is positive definite.
	\end{assumption}
	
	In light of the stability results presented in Section~\ref{sec:stability_single}, where Assumption~\ref{apt:positive_uh} is imposed on the plant, we extend a comparable assumption to the system $\widehat{\mathcal{H}}_{p}$. This parallels the conditions for the plant, giving a consistent framework for our output consensus results for networked systems.
	\begin{assumption}\label{apt:networked_positive_uh}
		For the system $\widehat{\mathcal{H}}_{p}$ with a constant input $\overline{\widehat{U}}_{p}$ which results in a constant output $\overline{\widehat{Y}}_{p}$, then $\overline{\widehat{U}}_{p}^{\top}\overline{\widehat{Y}}_{p} \geq 0.$ 
	\end{assumption}
	\medskip
	
	The following theorems establish output consensus results for the feedback system $(\widehat{\mathcal{H}}_{p},\mathcal{H}_{c})$.

		\begin{theorem}
			\label{thm:networked_feedback_static}
			Consider a nonlinear OSNI node plants $H_{pi}, i \in \mathcal{V}$. Suppose Assumptions~\ref{apt:state_output_consistency} and~\ref{apt:input_state_consistency} hold for each nonlinear OSNI node plant, and Assumption~\ref{apt:networked_positive_uh} holds for the networked node plants $\widehat{\mathcal{H}}_{p}$. Also consider nonlinear edge controllers $H_{cl}, l \in \mathcal{L}$, each of which is either a static controller or a dynamic NI controller. Suppose Assumptions~\ref{apt:g} and \ref{apt:negative_uy} hold for the nonlinear static controllers, and Assumptions~\ref{apt:g}, \ref{apt:state_output_consistency}, and \ref{apt:negative_uy} hold for the nonlinear dynamic NI controllers.   Further, consider the feedback system $(\widehat{\mathcal{H}}_{p}, \mathcal{H}_{c})$. Suppose  Assumption~\ref{apt:networked_local_domain} holds for the feedback system. Then, local output consensus is achieved.
		\end{theorem}

	{\it Proof.} First, according to Assumption~\ref{apt:networked_local_domain}, the candidate Lyapunov function~\eqref{eq:networked_Lyapunov} of the feedback system $(\widehat{\mathcal{H}}_{p}, \mathcal{H}_{c})$ is positive definite on an open domain $\mathcal D$. Second, we  analyze the time derivative of the candidate Lyapunov function~\eqref{eq:networked_Lyapunov}:
	\begin{align}
		\dot{\widehat{W}} 
		\overset{Eqs.~\eqref{eq:V_hat_OSNI}, \eqref{eq:NI_dot_V_c}}{\leq}  & \ (\widehat{U}_{p} -\Pi_{cx} -\Pi_{cu})^{\top}\dot{\widehat{Y}}_{p}  - \epsilon_{p\min}\|\dot{Y}_{p}\|^{2}  \notag\\
		&  + (U_{c} - Y_{p})^{\top}\dot{\Pi}_{cx} \notag\\
		\overset{Eq.~\eqref{eq:network_input_output}}{\leq} & - \epsilon_{p\min}\|\dot{Y}_{p}\|^{2} \leq  0, \label{eq:dot_W_hat}
	\end{align}
	where $\epsilon_{p\min} = \min\{\epsilon_{p1}, \dots, \epsilon_{pN}\} > 0.$  Therefore, the feedback system $(\widehat{\mathcal{H}}_{p}, \mathcal{H}_{c})$ is at least locally stable in the sense of Lyapunov. 
	
	 Equation~\eqref{eq:dot_W_hat} implies that $\dot{\widehat{W}}$ can remain zero only if $\dot{Y}_{p}$ remains zero. This implies the aggregated node plants $\mathcal{H}_{p}$ reach a steady state such that
		$
		\dot{Y}_{p} \equiv 0 \overset{\text{A}2}{\implies} \overline{Y}_{p} \overset{\text{A}2}{\implies} \overline{X}_{p}  \overset{\text{A}3}{\implies} \overline{U}_{p}.
		$
		The aggregated edge controllers $\mathcal{H}_{c}$ consists of  static controllers and dynamic NI controllers. The static controllers reach a steady state such that
		$	\overline{U}_{c}^{[s]} \overset{Eq.~\eqref{eq:Y_hat_p_U_c}}{\equiv} \overline{\widehat{Y}}_{p}^{[s]} \overset{Eq.~\eqref{eq:Y_hat_p_Y_p}}{\equiv} (\mathbf{Q}^{\top}Y_{p})^{[s]} \implies \overline{Y}_{c}^{[s]} \equiv \Pi_{cu}^{[s]}(\overline{U}_{c}^{[s]})$, where the supscript $[s]$ represents those static controllers. The dynamic controllers reach a steady state such that 
		$\overline{U}_{p} \equiv \overline{\mathbf{Q}Y_{c}} \text{ and } \overline{U}_{c} \equiv \overline{\widehat{Y}}_{p} \overset{Eq.~\eqref{eq:output_vector}}{\implies} \overline{Q \Pi_{cx}} \overset{\text{A}2}{\implies} \overline{X}_{c} \overset{\text{A}2, \text{A}3}{\implies} \overline{U}_{c}^{[d]} \text{ and } \overline{Y}_{c}^{[d]}$.
		Thus, we conclude that the networked node plants $\widehat{\mathcal{H}}_{p}$ are subject to the constant input $\overline{\widehat{U}}_{p}$ and have constant output $\overline{\widehat{Y}}_{p}$, where the aggregated edge controllers are subject to constant input $\overline{U}_{c}$ and constant output $\overline{Y}_{c}$. Then,
				$\overline{\widehat{U}}_{p}^{\top}\overline{\widehat{Y}}_{p} \equiv \overline{U}_{c}^{\top} \overline{Y}_{c},  \
				 \overline{\widehat{U}}_{p}^{\top}\overline{\widehat{Y}}_{p} \overset{\text{A}8}{\geq} 0, \text{ and }
				 \overline{U}_{c}^{\top}\overline{Y}_{c}  \overset{\text{A}5}{\leq} - \gamma_{c\min} \|\overline{U}_{c} \|^{2},$
		where $\gamma_{c\min} = \min\{\gamma_{c1},\dots,\gamma_{cL}\}.$
		This is only possible when $\overline{U}_{c} \equiv \overline{\widehat{Y}}_{p} \equiv 0$; i.e., achieving local output consensus. Therefore, we can conclude that $\dot{\widehat{W}}$ cannot remain zero unless $\overline{\widehat{Y}}_{p} = 0$. 
		The proof is now completed. \hfill $\square$

	\begin{theorem}
		\label{thm2:networked_feedback_NI/OSNI}
		Consider nonlinear NI node plants $H_{pi}, i \in \mathcal{V}$. Suppose Assumptions~\ref{apt:state_output_consistency} and~\ref{apt:input_state_consistency} hold for each nonlinear NI node plant, and Assumption~\ref{apt:networked_positive_uh} holds for the networked node plants $\widehat{\mathcal{H}}_{p}$. Also consider nonlinear OSNI edge controllers $H_{cl}, l \in \mathcal{L}$. Suppose Assumptions~\ref{apt:g}, \ref{apt:state_output_consistency},~\ref{apt:input_state_consistency}, and \ref{apt:negative_uy} hold for nonlinear dynamic OSNI controllers.   Further, consider the feedback system $(\widehat{\mathcal{H}}_{p}, \mathcal{H}_{c})$. Suppose  Assumption~\ref{apt:networked_local_domain} holds for the feedback system. Then, local output consensus is achieved.
	\end{theorem} 
	{\it Proof.} The proof is similar to that for Theorem~\ref{thm:networked_feedback_static}. \hfill$\square$

	\section{Application to Power Transmission Systems}\label{sec:power_application}
	In this section, we apply the proposed theoretical results to the practical problem of frequency synchronization and angle consensus in power transmission systems.
	
	\subsection{Transmission Network Model}
	Consider a transmission network comprised of $N$ nodes representing synchronous generator buses and $L$ edges representing transmission lines. The topology of the transmission network is represented by a connected and undirected graph $\mathcal{G} = (\mathcal{V},\mathcal{E})$. A generator bus has several components, including an AC generator, and fixed inflexible load. 
	The AC generator converts mechanical power into electric power through a rotating prime mover. The fixed inflexible load consumes a known amount of power. 
	
	For the transmission network, we denote the nominal frequency by $\omega^{0}$. For each generator bus $i \in \mathcal{V}$, we denote the rotor speed by $\omega_{i}$, and the rotor angle with respect to a rotating reference frame at the speed of $\omega^{0}$ by $\delta_{i}$. We adopt the following assumptions that are well-justified for power transmission systems \citep{kundur2022power}.
	\begin{enumerate}
		\item [(i)] Each transmission line $(i,j) \in \mathcal{E}$ is lossless and is thereby characterized by its reactance $ X_{ij}$.
		
		\item [(ii)] The internal voltage magnitude $E_{i}^{0}$ of each generator  bus $i \in \mathcal{V}$ is constant.
		
		\item [(iii)] Reactive power injections on buses are controlled to maintain an internal voltage magnitude $E_{i}^{0}$ for each generator bus $i \in \mathcal{V}$, where such controllers are decoupled from frequency and angle regulation in transmission grids.
	\end{enumerate}
	It is important to note that in this paper we do not conduct a small-signal stability analysis as our results are valid for both small and large rotor angle differences between generator buses. 
	
	{\bf Interconnected Swing Equations.}  Under assumptions (i)-(iii), the dynamics of each generator bus $i \in \mathcal{V}$ can be described by the following nonlinear swing equation \citep{dorfler2012synchronization}:
	\begin{subequations}\label{eq:swing_dynamics}
		\begin{align}
			\dot{\delta}_{i} &= \omega_{i} - \omega^{0},\\
			\dot{\omega}_{i} &= \frac{1}{M_{i}}\Big[P^{ M}_{i} - D_{i}(\omega_{i} - \omega^{0})  - P^{L}_{i} -  P^{ E}_{i}\Big].
		\end{align}
	\end{subequations}
	
	Here, $M_{i} > 0$ represents the inertia coefficient of generator bus $i$, and $D_{i} > 0$ represents the damping coefficient of generator bus $i$. The variables are defined as follows: $P^{ M}_{i}$ represents the mechanical power injection to generator bus $i$, $P^{L}_{i}$ represents the fixed power consumption of the load at generator bus $i$, and $P^{ E}_{i}$ represents the electric power output of generator bus $i$ to the transmission network. The electric power output of generator bus $i$ equals the net branch power flow from generator bus $i$ to other neighboring buses, represented by 
	$
	P^{ E}_{i} = \sum_{j \in \mathcal{N}(i)}P_{ij}.
	$
	The branch power flow $P_{ij}$ from generator bus $i$ to a neighboring bus $j \in \mathcal{N}(i)$ is characterized by
	\begin{equation}\label{eq:branch_flow}
		P_{ij} =  \frac{E_{i}^{0}E_{j}^{0}}{X_{ij}}\sin(\delta_{i} - \delta_{j}) = P_{ij}^{\max}\sin(\delta_{i} - \delta_{j}),
	\end{equation}
	where $X_{ij}$ is the reactance of the transmission line $(i,j)$, and $P_{ij}^{\max} = \frac{E_{i}^{0}E_{j}^{0}}{X_{ij}}$ is the maximum power transfer on the transmission line $(i,j)$.

	Combining Eqs.~\eqref{eq:swing_dynamics} and \eqref{eq:branch_flow}, the interconnected swing equations for the transmission network in terms of rotor angles $\delta_{i}$, for all $i \in \mathcal{V}$, are therefore given by 
	\begin{equation}\label{eq:delta_interconnected_swing}
		M_{i}\ddot{\delta_{i}} + D_{i}\dot{\delta_{i}} = P^{ M}_{i}  - P^{ L}_{i} - \sum_{j \in \mathcal{N}(i)}P_{ij}^{\max}\sin(\delta_{i} - \delta_{j}).
	\end{equation}
	
	In practice, faults arise in power systems that can cause a sudden change in the energy drawn through transmission lines (e.g., a power line falls to the ground resulting in a change in network impedance). Prior to a fault occurrence, the power transmission system~\eqref{eq:delta_interconnected_swing} is assumed to be at a stable equilibrium. Denote  the steady-state rotor angle difference between two neighboring buses $i$ and $j$ by $
	\overline{\psi}_{ij} = \overline{\delta}_{i} - \overline{\delta}_{j}.
	$
	Denote the stable equilibrium by
	\begin{equation}\label{eq:stable_equilibrium}
		(\overline{P}_{i}^{M}, \overline{P}_{i}^{L}, \overline{P}_{ij}^{\max}, \overline{\psi}_{ij}),  \forall i \in \mathcal{V},  \forall (i,j) \in \mathcal{E}.
	\end{equation}
	The relation between steady-state values is described by 
	\begin{equation}\label{eq:equilibrium}
		\overline{P}^{ M}_{i}  - \overline{P}^{ L}_{i} - \sum_{j \in \mathcal{N}(i)}\overline{P}_{ij}^{\max}\sin\overline{\psi}_{ij} = 0.
	\end{equation}
	In the aftermath of a fault, the bus frequencies $\omega_{i},\forall \in \mcalV$ deviate from their nominal frequency $\omega^{0}$ and the angle differences $\delta_{i} - \delta_{j}, \forall (i,j) \in \mcalE$ deviate from their steady-state values $\overline{\psi}_{ij}, \forall (i,j) \in \mcalE$.
	


	\subsection{Resilience in Power Transmission Systems}
	In what follows, we investigate transient conditions in the aftermath of fault, during which the mechanical power of synchronous generators, the load power consumption, and the maximum power transfer throughout the bulk grid, all return to pre-fault conditions. We also define the resilience of existing power transmission systems in the aftermath of a fault. Specifically, we consider the initial power angle deviation during transient conditions and asertain if it is in a suitable domain to enable the power transmission system to synchronize bus frequencies to the nominal and recover angle differences to their pre-fault values.
	
	For each generator bus $i \in \mcalV$, we denote the angle deviation from the pre-fault angle by $\widetilde{\delta}_{i}= \delta_{i} - \overline{\delta}_{i}$. For each transmission line $(i,j) \in \mathcal{E}$, we denote the angle deviation difference between the two connected generator buses by $\widetilde{\psi}_{ij}=\widetilde{\delta}_{i} - \widetilde{\delta}_{j}$. Using Eq.~\eqref{eq:delta_interconnected_swing} and Eq.~\eqref{eq:equilibrium}, the interconnected swing equations can be specifically rewritten in terms of rotor angle deviations $\widetilde{\delta}_{i}, i \in \mathcal{V}$:
	\begin{equation}\label{eq:swing_deviation}
		\begin{aligned}
			M_{i}\ddot{\widetilde{\delta}}_{i} + D_{i}\dot{\widetilde{\delta}}_{i}  = \sum_{j \in \mcalN(i)}\overline{P}_{ij}^{\max}\big(\sin\overline{\psi}_{ij} - \sin(\widetilde{\psi}_{ij} + \overline{\psi}_{ij})\big),
		\end{aligned}
	\end{equation}
	with initial angle deviations $\widetilde{\delta}_{i}(0), \forall i \in \mathcal{V}.$ 
	
	By reformulating the system~\eqref{eq:swing_deviation} into the feedback interconnection of  OSNI  node plants and static edge controllers based on the underlying transmission network, we can utilize the result of output consensus as established in Section~\ref{sec:networked_case} to prove existing power transmission systems can synchronize bus frequencies and recover angle differences when the initial deviation is in a suitable domain. 
	
	Define $x_{pi} = [\dot{\widetilde{\delta}}_{pi},  \widetilde{\delta}_{pi} ]^{\top} \in \mathbb{R}^{2}$, $u_{pi} \in \mathbb{R}$, $x_{cl} \equiv  0$, and $u_{cl} = \widetilde{\psi}_{ij} \in \mathbb{R}$ with $j \in \mathcal{N}(i)$. The system \eqref{eq:swing_deviation} can be represented by the interconnection of the node systems and the edge systems based on the transmission network:
	\begin{subequations}
		\label{sys:reformulate_with_damp}
		\begin{align}
			H_{pi} \ : \quad	\dot{x}_{pi} &= A_{pi}x_{pi} + B_{pi}u_{pi}, \label{eq:hpi_x}\\
			y_{pi} & = C_{pi}x_{pi}, \label{eq:hpi_y} \\
			H_{cl} \ :  \quad y_{cl} & = \overline{P}^{\max}_{l}\big( \sin\overline{\psi}_{l} -  \sin(u_{cl} + \overline{\psi}_{l}) \big), \label{eq:hcl_y}
		\end{align}
	\end{subequations}
	for all $i \in \mathcal{V}$ and all $l \in \mathcal{L}$. The system $H_{pi}$  has system matrices 
	$$
	A_{pi} = \begin{bmatrix}
		\frac{-D_{i}}{M_{i}} & 0 \\
		1 & 0 
	\end{bmatrix}, \ B_{pi} = \begin{bmatrix}
		\frac{1}{M_{i}} \\
		0
	\end{bmatrix}, \text{ and } \ C_{pi} = \begin{bmatrix}
		0 & 1
	\end{bmatrix}.
	$$
	The system $H_{cl}$ has system parameters $$\overline{P}^{\max}_{l} = \overline{P}^{\max}_{ij} \text{ and } \sin\overline{\psi}_{l} = \sin\overline{\psi}_{ij},$$ when the node $i$ and the node $j$ are the initial node and the terminal node, respectively.
	The input-output relation induced from the feedback interconnection based on the transmission network is such that 
	\begin{align}\label{eq:transmission_feedback_relation}
		u_{pi} =  \sum_{l \in \mathcal{L}} q_{il}y_{cl} \ \text{ and }  \ u_{cl} = \sum_{k \in \mathcal{V}}q_{kl}y_{pk}. 
	\end{align}
	
	\begin{remark}
		The system~\eqref{eq:swing_deviation} does not inherently incorporate controllers. For the sake of applying our results established in Section~\ref{sec:networked_case} more directly, we designate the left side of Eq.~\eqref{eq:swing_deviation} as ``node plants'' corresponding to actual dynamics in generator buses and the right side as ``edge controllers'' corresponding to power flows on transmission lines. Furthermore, this choice provides insights into how we can leverage batteries to create virtual transmission lines, thereby enhancing the robustness margin of the power transmission systems in Section~\ref {sec:battery-based controllers}.
	\end{remark}
	
	In the following theorem, we prove that  the power transmission system can achieve local output consensus, thereby regulating bus frequencies and restoring power angle differences. Denote a local domain by $\mathcal{D} = \mathcal{D}_{1} \cap \mathcal{D}_{2},$ where
	\begin{align*}
		\mathcal{D}_{1} = \{\widetilde{\delta}_{i}, & \forall i \in \mathcal{V} \ | \ \widetilde{\psi}_{l} \in (-\pi -2 \overline{\psi}_{l}, \pi -2 \overline{\psi}_{l}), \forall l \in \mathcal{L}\}, \\
		\mathcal{D}_{2} =  \bigl\{\widetilde{\delta}_{i}, &\forall i \in \mathcal{V} \ | \ \sum_{l \in \mathcal{L}} \overline{P}_{l}^{\max}\big(\cos\overline{\psi}_{l} - \widetilde{\psi}_{l}\sin\overline{\psi}_{l} \notag \\ &-\cos(\widetilde{\psi}_{l}+\overline{\psi}_{l}) \big) > 0\bigr\} \cup \{\widetilde{\delta}_{i} = 0, \forall i \in \mathcal{V}\}. 
	\end{align*}
	
	\begin{theorem}\label{thm:power_systems_no_controller}
		Consider node plants $\mathcal{H}_{p}$ described by~\eqref{eq:hpi_x}-\eqref{eq:hpi_y} and edge controllers $\mathcal{H}_{c}$ described by~\eqref{eq:hcl_y}. Consider the feedback interconnection of node plants and edge controllers based on the underlying transmission network as depicted in Fig.~\ref{fig:networked_sys}. Then, the feedback system achieves local output consensus.
	\end{theorem}
	{\it Proof.} First, since each node plant is a linear system, Assumptions~\ref{apt:state_output_consistency} and ~\ref{apt:input_state_consistency} are satisfied.  We are to show that each node plant \eqref{eq:hpi_x}-\eqref{eq:hpi_y} is OSNI. We choose the storage function of  each node plant  as 
	\begin{align*}
		V_{pi} = x_{pi}^{\top}P_{pi}x_{pi} =  [\dot{\widetilde{\delta}}_{i} \ \widetilde{\delta}_{i}]  \begin{bmatrix}
			\frac{M_{i}}{2} & 0\\
			0 & 0
		\end{bmatrix} [\dot{\widetilde{\delta}}_{i} \ \widetilde{\delta}_{i}]^{\top} = \frac{M_{i}}{2}\dot{\widetilde{\delta}}_{i}^{2}.
	\end{align*}
	We analyze the time derivative of the storage function of  each node plant:
	\begin{align*}
		\dot{V}_{pi} =  \dot{\widetilde{\delta}}_{i} u_{pi} -D_{i}\dot{\widetilde{\delta}}_{i}^{2} \leq u_{pi}^{\top}\dot h_{pi}(x_{pi}) - \epsilon_{pi}\|\dot{h}_{pi}(x_{pi}) \|^{2}.
	\end{align*}
	Thus, with $0 < \epsilon_{pi} \leq D_{i}$, each node plant satisfies the definition of OSNI systems. Furthermore, according to Lemma~\ref{lemma:post-processed}, the networked node plants $\widehat{\mathcal{H}}_{p}$ is at least linear NI. Every linear NI system satisfies Assumption~\ref{apt:networked_positive_uh} \citep{shi2023output}. Second, for static edge controllers, we can verify that Assumption~\ref{apt:g} is satisfied; i.e., $y_{cl}(0) = 0$, and also validate that Assumption~\ref{apt:negative_uy} is satisfied in the domain $\mathcal{D}_{1}.$ Third, for the feedback system $(\mathcal{H}_{p}, \mathcal{H}_{c})$, we can validate Assumption~\ref{apt:local_domain}, where the candidate Lyapunov function~\eqref{eq:networked_Lyapunov} is  positive definite in the domain $\mathcal{D}_{2}$. Therefore, Theorem~\ref{thm:networked_feedback_static} can be applied and the feedback system achieves local consensus. The proof is completed. \hfill$\square$
	
	\subsection{Battery-based Angle Feedback Controllers}\label{sec:battery-based controllers}
	In what follows, we equip generator buses with large-scale batteries and we design angle based feedback controllers for the battery-based actuators. We design the angle based feedback controllers to enhance the transient stability of the power transmission system.
	
	We consider equipping batteries at generator buses. Hence, modifying Eq.~\eqref{eq:delta_interconnected_swing}, the swing equation is revised as 
	\begin{equation}\label{eq:storage_swing}
		M_{i}\ddot{\delta_{i}} + D_{i}\dot{\delta_{i}} = P^{ M}_{i} + P^{ST}_{i} - P^{ L}_{i} - P^{E}_{i},
	\end{equation}
	where $P^{ ST}_{i}$ represents the power output of the battery at each generator bus $i \in \mathcal{V}$.  
	
	Prior to a fault occurrence, the system~\eqref{eq:storage_swing} is at a stable equilibrium 
	$		(\overline{P}_{i}^{M}, \overline{P}_{i}^{ST}, \overline{P}_{i}^{L}, \overline{P}_{ij}^{\max}, \overline{\psi}_{ij}),  \forall i \in \mathcal{V},  \forall (i,j) \in \mathcal{E},$
	where
	$		\overline{P}^{ M}_{i} + \overline{P}_{i}^{ST} - \overline{P}^{ L}_{i} - \sum_{j \in \mathcal{N}(i)}\overline{P}_{ij}^{\max}\sin\overline{\psi}_{ij} = 0.$ 
	In the aftermath of a fault, the generator bus frequency deviates from its nominal value and the angle differences deviate from pre-fault conditions. In what follows, we investigate the post-fault transients, during which the mechanical power of synchronous generators and the maximum power transfer of the transmission network return back to pre-fault values.

	We define the change in storage power output, $\widetilde{P}^{ST}_{i} = P^{ST}_{i} - \overline{P}^{ST}_{i}$, as the difference from the storage power output before a fault occurs. The swing equation~\eqref{eq:storage_swing} can be rewritten in terms of the angle deviation $\tilde{\delta}_{i}, i \in \mathcal{V}$ specifically:
	\begin{equation*}\label{eq:storage_deviation_damp}
		M_{i}\ddot{\widetilde{\delta}}_{i} + D_{i}\dot{\widetilde{\delta}}_{i}  = \widetilde{P}^{ST}_{i} + \sum_{j \in \mcalN(i)}\overline{P}_{ij}^{\max}\big(\sin\overline{\psi}_{ij} - \sin(\widetilde{\psi}_{ij} + \overline{\psi}_{ij})\big).
	\end{equation*}
	Without battery based controllers, existing power transmission systems can synchronize bus frequencies and recover angle differences when the initial deviation is in a suitable domain. By employing battery-based controllers as  NI controllers, the transient stability of power transmission systems can also be guaranteed.
	
In what follows,  we investigate a way in which the power grid can be gradually transitioned into an NI controlled system, one transmission line at a time. We consider co-locating batteries with generators, positioned at both ends of the $k$th transmission line. While the edge controllers $l \neq k$ are described by Eq.~\eqref{eq:hcl_y},  the edge controller $k$ is designed as
		\begin{subequations}
			\label{sys:control}
			\begin{align}
				H_{ck}: \quad	\dot{x}_{ck} &= 
				-\frac{1}{\tau}x_{ck} + \frac{K_{1}}{\tau}u_{ck}, \\
				y_{ck} 
				&= x_{ck} - K_{2}u_{ck}, 
			\end{align}
		\end{subequations}
		where $\tau >0$ and $K_{2} > K_{1} >0$. 
		
		\begin{theorem}
			Consider node plants $\mathcal{H}_{p}$ described by~\eqref{eq:hpi_x}-\eqref{eq:hpi_y}, edge controllers $l \neq k$ described by the system \eqref{eq:hcl_y}, and edge controller $k$ described by the system \eqref{sys:control}. Consider the feedback interconnection of node plants and edge controllers based on the underlying transmission network as depicted in Fig.~\ref{fig:networked_sys}. Then, the feedback system achieves local output consensus.
		\end{theorem}
  {\it Proof.} In the proof for Theorem~\ref{thm:power_systems_no_controller}, we have demonstrated that each node plant is OSNI and fulfills Assumptions~\ref{apt:state_output_consistency} and ~\ref{apt:input_state_consistency}; furthermore, the networked node plants satisfy Assumption~\ref{apt:networked_positive_uh}. For edge controllers $l \neq k$, we choose the storage function as $V_{cl} \equiv 0, l \neq k$. For the edge controller $k$, we choose the storage function as $V_{ck}(x_{ck}) = \frac{x_{ck}^{2}}{2K_{1}}$. We can verify that $\dot{V}_{ck}(x_{ck}) = \frac{1}{K_{1}}x_{ck}\dot{x}_{ck} = \frac{1}{K_{1}}(-\tau \dot{x}_{ck} + K_{1}u_{ck})\dot{x}_{ck} \leq u_{ck}^{\top}\dot{h}_{ck}(x_{ck}).$ Hence, the edge controller $k$ is NI. Furthermore, a constant input $\overline{u}_{ck}$ results in a constant state $\overline{x}_{ck} = K_{1}\overline{u}_{ck}$ and a constant output $\overline{y}_{ck} = (K_{1} - K_{2})\overline{u}_{ck}.$ We can also verify Assumption~\ref{apt:negative_uy} that $\overline{u}_{ck}^{\top}\overline{y}_{ck} = (K_{1} - K_{2})\overline{u}_{ck}^{2} \leq -\gamma_{ck} \overline{u}_{ck}^{2} $, where $\gamma_{ck} \leq K_{2} - K_{1}$. For the feedback system, we look at the candidate Lyapunov function~\eqref{eq:networked_Lyapunov} is positive definite in the domain 
  \begin{align*}
		\mathcal{D} =  \bigl\{\widetilde{\delta}_{i}, &\forall i \in \mathcal{V} \ | \ \sum_{l \neq k} \overline{P}_{l}^{\max}\big(\cos\overline{\psi}_{l} - \widetilde{\psi}_{l}\sin\overline{\psi}_{l} \notag \\ &-\cos(\widetilde{\psi}_{l}+\overline{\psi}_{l}) \big) > 0\bigr\} \cup \{\widetilde{\delta}_{i} = 0, \forall i \in \mathcal{V}\}. 
	\end{align*}
    Therefore, Theorem~\ref{thm:networked_feedback_static} can be applied and the feedback system achieves local consensus.\hfill$\square$
        
    \begin{remark}
       The edge controller $k$ can equivalently be viewed as two end node controllers operating in a distributed manner. The control action actuated by the battery at the end generator bus $i$ is described by
    $$
    \widetilde{P}^{ST}_{i} = q_{ik}\big(x_{ck} - K_{2}u_{ck} - \overline{P}^{\max}_{l}( \sin\overline{\psi}_{k} -  \sin(u_{ck} + \overline{\psi}_{k}) )\big),
    $$
    where $q_{ik}$ is the $k$th element in the $i$th row of the incidence matrix $\mathbf{Q}$.
    \end{remark}

	\section{Conclusion}\label{sec:conclusion}
	In this paper, we proposed a networked control framework for power transmission systems. A novel Lur'e-Postnikov-like Lyapunov function was formulated, and stability proofs were constructed for the feedback interconnection of two single nonlinear NI systems. Additionally, output feedback consensus results were established for the feedback interconnection of two networked nonlinear NI systems based on the network topology. Our theoretical framework provided support for the design of battery-based control in power transmission systems. We demonstrated a way in which the electric power grid could be gradually transitioned into the proposed NI controlled system, one transmission line at a time. 
 A possible future research direction will include saturation of battery actuators for networked power transmission systems.
	\bibliography{ifacconf}
	
\end{document}